\documentclass[onecolumn]{aastex631}
\usepackage{multirow}
\usepackage{enumerate}
\usepackage{diagbox}
\usepackage{enumitem}
\usepackage{amssymb}
\usepackage{amsmath}

\shorttitle{Revised Constraints on the FRB population}
\shortauthors{H.-N. Lin and R. Zou}
\graphicspath{{./}{figures/}}

\begin{document}

\title{Revised Constraints on the fast radio burst population from the first CHIME/FRB catalog}

\correspondingauthor{Rui Zou}\email{zourui@stu.cqu.edu.cn}

\author[0000-0003-1659-3368]{Hai-Nan Lin}
\affiliation{Department of Physics, Chongqing University, Chongqing 401331, China}
\affiliation{Chongqing Key Laboratory for Strongly Coupled Physics, Chongqing University, Chongqing 401331, China}

\author[0000-0002-9998-6222]{Rui Zou}
\affiliation{Department of Physics, Chongqing University, Chongqing 401331, China}
\affiliation{Chongqing Key Laboratory for Strongly Coupled Physics, Chongqing University, Chongqing 401331, China}

\begin{abstract}
In this paper, we investigate the FRB population using the first CHIME/FRB catalog. We first reconstruct the extragalactic dispersion measure -- redshift relation ($\mathrm{DM_E} - z$ relation) from well-localized FRBs, then use it to infer redshift and isotropic energy of the first CHIME/FRB catalog. The intrinsic energy distribution is modeled by the power law with an exponential cutoff, and the selection effect of the CHIME telescope is modeled by a two-parametric function of specific fluence. For the intrinsic redshift distribution, the star formation history (SFH) model, as well as other five SFH-related models are considered. We construct the joint likelihood of fluence, energy and redshift, and all the free parameters are constrained simultaneously using Bayesian inference method. The Bayesian information criterion (BIC) is used to choose the model that best matches the observational data. For comparison, we fit our models with two data samples, i.e. the Full sample and the Gold sample. The power-law index and cutoff energy are tightly constrained to be $1.8 \lesssim \alpha \lesssim 1.9$ and $\mathrm{log}(E_c/{\rm erg}) \approx 42$, which are almost independent of the redshift distribution model and the data sample we choose. The parameters involving the selection effect strongly depends on the data sample, but are insensitive to the redshift distribution model. According to BIC, the pure SFH model is strongly disfavored by both the Full sample and Gold sample. For the rest five SFH-related redshift distribution models, most of them can match the data well if the parameters are properly chosen. Therefore, with the present data, it is still premature to draw a conclusive conclusion on the FRB population.
\end{abstract}

\section{Introduction}\label{sec:introduction}

Fast radio bursts (FRBs) are very bright radio pulses with duration of milliseconds that randomly happen in the sky \citep{Petroff:2019tty,Cordes:2019cmq,Zhang:2020qgp,Xiao:2021omr}. The first FRB was declared in 2007, when \citet{Lorimer:2007qn} re-examined the 2001 archive data of the Parkes 64-m telescope in Australia. Till now, hundreds of FRBs have been detected by different telescopes \citep{Petroff:2016tcr,CHIMEFRB:2021srp}, such as the Canadian Hydrogen Intensity Mapping Experiment \cite[CHIME,][]{Amiri_2018} and the Five-hundred-meter Aperture Spherical Telescope \cite[FAST,][]{Nan:2011um}, etc. Their large values of dispersion measure (DM) in excess of the contribution of the Milky Way indicate an extragalactic or even cosmological origin \citep{Keane:2016yyk,Chatterjee:2017dqg,Tendulkar:2017vuq}, except one that is verified to be generated from the Milky Way \citep{CHIMEFRB:2020abu,Bochenek:2020zxn}. FRBs are generally classified into repeating and non-repeating. Most of the sources observed by the CHIME telescope are apparently non-repeating. The first repeating FRB 121102 was discovered by the Arecibo radio telescope \citep{Spitler:2016dmz,Scholz:2016rpt}, and soon its host galaxy was identified and the cosmological origin was confirmed \citep{Chatterjee:2017dqg,Marcote:2017wan,Tendulkar:2017vuq}. Since then, our knowledge on FRBs has significantly advanced.

A mystery that still puzzles scientists is the redshift distribution of FRBs, which is highly correlated with their progenitors, i.e, different origins produce different redshift distributions \citep{Zhang:2020ass}. For instance, if FRBs originate from magnetars, it's reasonable to assume that they follow the cosmic star formation history (SFH) \citep{James:2021oep,Zhang:2020ass,CHIMEFRB:2021srp}, because galactic magnetars are believed to come from massive star deaths \citep{Kaspi:2017fwg}. Recently, the landmark identification of Galactic FRB 20200428A, together with an X-ray burst from the magnetar SGR 1935+2154 prove the magnetar origin of some, if not all FRBs \citep{Andersen:2020hvz,Bochenek:2020zxn,Insight-HXMTTeam:2020dmu,Mereghetti:2020unm,Tavani:2020adq,Ridnaia:2020gcv}. The hypothesis that most FRBs track SFH was further confirmed by the Parkes and ASKAP FRB samples \citep{Zhang:2020ass,James:2021oep}. However, the SFH model fails to explain the large sample of CHIME/FRB \citep{Zhang:2021kdu,Qiang:2021ljr}. The discovery of the repeating source FRB 20200120E situated in a globular cluster of the nearby galaxy M81 hints that some FRBs may track old stellar populations \citep{Bhardwaj:2021xaa,Kirsten:2021llv,Nimmo:2021yob}. In addition, delayed models that FRBs track the compact binary mergers are also acceptable \citep{Cao:2018yzp,Locatelli:2018anc,Zhang:2020ass,Zhang:2021kdu}. Above all, the intrinsic redshift distribution of FRBs is still in extensive debate.

Though the number of detected FRBs is rapidly increasing due to the development of observation technique, the sample that has direct redshift measurement is still small \citep{Heintz_2020}. Several works use DM as a rough proxy of redshift, i.e., ${\mathrm{DM}\approx 1000z{\rm ~pc~cm^{-3}}}$ \citep{Lorimer:2018rwi}, as confirmed by theoretical analyses and multiple observations \citep{Ioka:2003fr,Inoue:2003ga,Deng:2013aga,Macquart:2020lln}. However, due to the complex components of DM, this formula is very rough and may cause significant bias if it is used to estimate redshift. For example, although the DM of FRB 20190520B is as large as $1200~{\rm pc~cm^{-3}}$, the redshift is measured to be only $z\sim 0.24$ \citep{Niu:2021bnl}. A more reasonably way is to reconstruct the ${\rm DM}-z$ relation (the so-called Macquart relation) by properly taking the probability distributions of the DMs of intergalactic medium (IGM) and host galaxy into account, and use the Macquart relation to infer redshift \citep{Macquart:2020lln}.

Except for the redshift distribution, the intrinsic energy distribution is crucial for studying FRB population as well. The intrinsic energy distribution is well-constrained regardless of the redshift distribution. It is commonly believed to follow the power law $dN/dE \propto  E^{-\alpha}$, with $1.8 \lesssim \alpha \lesssim 2.0$ \citep{Lu:2019pdn,Lu:2020nsg,Luo:2018tiy,Luo:2020wfx}. \citet{Luo:2020wfx} and \citet{Lu:2020nsg} argued that there may be a high-energy exponential cutoff $E_c$, which is estimated to be around $\mathrm{log}(E_c/{\rm erg}) \approx 41.5$. The cutoff energy is still not tightly constrained yet, and other values are also acceptable \citep{Zhang:2020ass}. Note that the specific fluence is a convolution of isotropic energy and redshift. It should follow a simple law $N \propto F_\nu ^{-3/2}$, and is insensitive to either redshift distribution or energy distribution \citep{Zhang:2021kdu}. The recently published CHIME/FRB catalog provides a relatively large sample for one to study the FRB population \citep{CHIMEFRB:2021srp}. After considering the selection effect of observational instrument, a properly chosen redshift distribution model should fit the data well. The aim of this paper is to choose a redshift distribution model that can best match the observational data.

In this paper, we first use the reconstructed $\mathrm{DM_E}-z$ relation to infer the redshift of CHIME/FRBs. Then the SFH model, along with five SFH-related models proposed by \citet{Qiang:2021ljr} are adopted to describe the possible redshift distribution. After considering the selection effect of telescope, we construct the distributions of fluence, energy and redshift. Finally we use the Beyesian inference method to constrain the model parameters by fitting the models to the CHIME/FRBs. The rest parts of this paper are organized as follows: In Section \ref{sec:host_frbs}, we reconstruct the $\mathrm{DM_E}-z$ relation from well-localized FRBs, then use it to infer the redshift and calculate the energy of CHIME/FRBs. In Section \ref{sec:population}, we consider six redshift distribution models, and construct a Bayesian framework to constrain the parameters. Finally, discussion and conclusions are drawn in Section \ref{sec:conclusions}.

\section{Inferring the redshift of CHIME/FRB}\label{sec:host_frbs}

The speed of electromagnetic waves travelling through cosmic plasma is frequency-dependent, that is, high-frequency electromagnetic waves travel faster than the low-frequency ones. The relative delay of arrival time is proportional to the dispersion measure (DM), which is defined as the integral of free electron number density along the line-of-sight. The observed DM of an extragalactic FRB can be decomposed into four terms \citep{Deng:2013aga,Gao:2014iva,Macquart:2020lln},
\begin{equation}\label{eq:DM_obs}
  {\rm DM_{obs}}={\rm DM_{MW,ISM}}+{\rm DM_{MW,halo}}+{\rm DM_{IGM}}+\frac{{\rm DM_{host}}}{1+z},
\end{equation}
where the subscripts ``$\mathrm{MW,ISM}$'', ``$\mathrm{MW,halo}$'', ``$\mathrm{IGM}$'', ``$\mathrm{host}$'' denote contributions from the Milky Way interstellar medium (ISM), the Galactic halo, the intergalactic medium (IGM) and the FRB host galaxy, respectively. Notice that the $\mathrm{DM_{host}}$ is in the source frame, which is weighted by $(1+z)^{-1}$ to transform it to the observer frame, with $z$ being the source redshift.

The Milky Way ISM term ($\mathrm {DM_{MW,ISM}}$) can be estimated using the Milky Way electron density models, such as the NE2001 model \citep{Cordes:2002wz} and the YMW16 model \citep{Yao_2017msh}. These two models predict consistent results for FRBs at high Galactic latitude, but the YMW16 model may overestimate the $\mathrm {DM_{MW,ISM}}$ value for FRBs at low Galactic latitude \citep{KochOcker:2021fia}. Therefore, we use the NE2001 model to calculate $\mathrm {DM_{MW,ISM}}$ in this work. The Galactic halo term is estimated to be about $\mathrm {DM_{MW,halo}} \approx 50\ \mathrm{pc~cm^{-3}}$, although the exact value of this term is poorly constrained yet \citep{Prochaska:2019mkd, Macquart:2020lln}. Here we follow \citet{Macquart:2020lln} to fix $\mathrm {DM_{MW,halo}} = 50 \ \mathrm{pc~cm^{-3}}$. We emphasise that the concrete value of $\mathrm {DM_{MW,halo}}$ shouldn't strongly bias our analyses, since its uncertainty is much smaller than that of the last two terms, i.e., $\mathrm{DM_{IGM}}$ and $\mathrm{DM_{host}}$, which we will be discussed in detail below.

By subtracting the first two terms from the total $\mathrm{DM_{obs}}$, we define the extragalactic DM as
\begin{equation}\label{eq:DM_E}
  {\rm DM_E}\equiv {\rm DM_{obs}}-{\rm DM_{MW,ISM}}-{\rm DM_{MW,halo}}={\rm DM_{IGM}}+\frac{{\rm DM_{host}}}{1+z}.
\end{equation}
Only the term $\mathrm {DM_{IGM}}$ contains cosmological information, the expression of which in the flat $\Lambda \mathrm{CDM}$ model reads \citep{Deng:2013aga,Zhang:2020ass}
\begin{equation}\label{DM_IGM_average}
  \langle{\rm DM_{IGM}}(z)\rangle=\frac{3cH_0\Omega_bf_{\rm IGM}f_e}{8\pi Gm_p}\int_0^z\frac{1+z}{\sqrt{\Omega_m(1+z)^3+\Omega_\Lambda}}dz,
\end{equation}
with $c$ the speed of light, $m_p$ the proton mass, $G$ the Newtonian gravitational constant. $f_e = Y_{\rm H}X_{e,{\rm H}}(z) + \frac{1}{2}Y_{\rm He}X_{e,\rm He}(z)$ is the electron fraction, denoting the extent of ionization progress of hydrogen and helium. $Y_\mathrm{H}$ and $Y_\mathrm{He}$ are the cosmological mass fractions of hydrogen and helium, which are estimated to be around $\sim 0.75$ and $\sim 0.25$, respectively. Given the electrons in both hydrogen and helium are fully ionized at $z<3$ \citep{Meiksin:2007rz,Becker:2010cu}, we take the ionization fractions to be $X_{e,\rm H} = X_{e,\rm He} = 1$. The Hubble constant $H_0$, the baryon mass density $\Omega_b$, the matter density $\Omega_m$ and the vacuum energy density $\Omega_\Lambda$ are all fixed to the Planck 2018 parameters, i.e., $H_0=67.4~{\rm km~s^{-1}~Mpc^{-1}}$, $\Omega_m=0.315$, $\Omega_\Lambda=0.685$ and $\Omega_{b}=0.0493$ \citep{Planck:2018vyg}. As for the cosmic baryon mass fraction $f_\mathrm{IGM}$, we know little about the evolution model of this term at present. Several observations indicate that $f_\mathrm{IGM}$ is about $0.84$ \citep{Fukugita:1997bi, Inoue:2003ga}. With five well-localized FRBs detected by ASKAP, \citet{Li:2020qei} also drew the similar conclusion. Therefore, we fix $f_\mathrm{IGM} = 0.84$ throughout this paper.

Though the average value of $\mathrm{DM_{IGM}}$ is given by equation (\ref{DM_IGM_average}), the real value of which should oscillate around the mean owing to the large-scale matter density fluctuation. Theoretical analyses as well as numerical simulations show that  the distribution of $\mathrm{DM_{IGM}}$ can be modeled as a long-tailed quasi-Gaussian distribution \citep{McQuinn:2013tmc,Macquart:2020lln,Zhang:2020xoc},
\begin{equation}\label{eq:P_delta}
  p_{\rm IGM}(\Delta)=A\Delta^{-\beta}\exp\left[-\frac{(\Delta^{-\alpha}-C_0)^2}{2\alpha^2\sigma_{\rm IGM}^2}\right], ~~~\Delta>0,
\end{equation}
where $\Delta\equiv{\rm DM_{IGM}}/\langle{\rm DM_{IGM}}\rangle$, and $\sigma _\mathrm{IGM} = Fz^{-0.5}$ denotes the effective standard deviation, with $F$ a free parameter that quantifies the strenth of baryon feedback \citep{Macquart:2020lln}. $\alpha$ and $\beta$ are two parameters related to the inner density profile of gas in haloes, which are chosen to be $\alpha = \beta = 3$ \citep{Macquart:2020lln}. $C_0$ and $A$ are constants dependent on $\sigma _\mathrm{IGM}$, where $C_0$ is chosen to guarantee that the mean value of $\Delta$ is unity, and $A$ to normalize the distribution accordingly.

At present, there is still few knowledge on the host term $\mathrm{DM_{host}}$ due to the absence of detailed observation on the local environment of most FRB sources. It may vary in a large scale, ranging from several tens \citep{Xu:2021qdn} to several hundreds \citep{Niu:2021bnl} ${\rm pc~cm^{-3}}$. To account for the large variation of $\mathrm{DM_{host}}$, it is usually modeled by the log-normal distribution \citep{Macquart:2020lln,Zhang:2020mgq},
\begin{equation}\label{eq:P_host}
  p_{\rm host}({\rm DM_{host}}|\mu,\sigma_{\rm host})=\frac{1}{\sqrt{2\pi}{\rm DM_{host}}\sigma_{\rm host}} \exp\left[-\frac{(\ln {\rm DM_{host}}-\mu)^2}{2\sigma_{\rm host}^2}\right],
\end{equation}
with $\mu$ and $\sigma_{\rm host}$ the mean and standard deviation of $\ln {\rm DM_{host}}$, respectively. At present, there is no strong evidence for the redshift-dependence of $\mu$ or $\sigma_{\rm host}$ \citep{Tang:2023qbg}. Thus we treat these two parameters as constants.

After properly taking the distributions of $\mathrm{DM_{IGM}}$ and $\mathrm{DM_{host}}$ into account, one can construct the distribution of ${\rm DM_E}$ at redshift $z$ by marginalizing over either term as \citep{Macquart:2020lln}
\begin{equation}\label{eq:P_E}
  p_E({\rm DM_E}|z)=\int_0^{(1+z)\rm DM_E}p_{\rm host}({\rm DM_{host}}|\mu,\sigma_{\rm host})p_{\rm IGM}({\rm DM_E}-\frac{\rm DM_{host}}{1+z}|F,z)d{\rm DM_{host}},
\end{equation}
where ($F,\mu,\sigma_{\rm host}$) are free parameters which can be constrained by fitting to a sample of well-localized FRBs.

Using 17 well-localized FRBs, \citet{Tang:2023qbg} get the parameters($F,\mu,\sigma_{\rm host}$) well constrained. By fixing the parameters to the best-fitting values, equation (\ref{eq:P_E}) can be used to reconstruct the $\mathrm{DM_E}-z$ relation, see Figure 2 of \citet{Tang:2023qbg}. Then the $\mathrm{DM_E}-z$ relation can be used to infer the redshift of FRBs that have no direct measurement of redshift but have well determined $\mathrm{DM_E}$, e.g. the first CHIME/FRB catalog.

The first CHIME/FRB catalog consists of 536 bursts, with 474 apparently non-repeating bursts and 62 repeating bursts from 18 FRB sources \citep{CHIMEFRB:2021srp}. All of the bursts have well detected ${\rm DM_{obs}}$, but most of them are absent of redshift measurement. We calculate ${\rm DM_{E}}$ of the 474 apparently non-repeating FRBs by subtracting ${\rm DM_{MW,ISM}}$ and ${\rm DM_{MW,halo}}$ from the total ${\rm DM_{obs}}$. Then the $\mathrm{DM_E}-z$ relation is used to calculated the redshift of FRBs with $\mathrm{DM_E}>100\ \mathrm{pc~cm^{-3}}$  (444 FRBs in total). In terms of FRBs with $\mathrm{DM_E}<100\ \mathrm{pc~cm^{-3}}$, the $\mathrm{DM_{host}}$ term dominates over $\mathrm{DM_{IGM}}$, which may cause large uncertainty on the redshift estimation. The inferred redshifts are listed in the {\it online materials} of \citet{Tang:2023qbg}. The inferred redshifts are in the range of $z_{\rm inf}\in(0.023,3.935)$. Only three bursts have inferred redshift larger than 3, i.e., FRB20180906B with $z_{\rm inf}=3.935_{-0.705}^{+0.463}$, FRB20181203C with $z_{\rm inf}=3.003_{-0.657}^{+0.443}$, and FRB20190430B with $z_{\rm inf}=3.278_{-0.650}^{+0.449}$. For the sake of statistical significance, we omit the three sparse data points at redshift $z>3$ in the following FRB population analyses.

With the observed fluence listed in the CHIME/FRB catalog and the inferred redshift, the isotropic energy of an FRB can be calculated as \citep{James:2021jbo}
\begin{equation}\label{eq:energy}
  E = \frac{4\pi d_\mathrm{L}^2}{(1+z)^{2+\beta}}F_\nu \Delta\nu
\end{equation}
where $d_\mathrm{L}$ is the luminosity distance, $\beta$ is the spectrum index ($F_\nu \propto \nu^\beta$), and $\Delta\nu$ is the bandwidth in which FRB is detected. \citet{Macquart:2018rsa} showed that $\beta = -1.5$ performs a good fit based on a sample of ASKAP/FRBs, hence we fix $\beta = -1.5$ in this work. In the first CHIME/FRB catalog, the observed fluence is averaged in the range $400\sim 800 \mathrm{MHz}$, indicating $\Delta\nu = 400 \mathrm{MHz}$. The 436 FRBs with fluence, energy and redshift are called the Full sample and are collected in the \textit{online material} of \citet{Tang:2023qbg}.

\begin{figure}[htbp]
  \centering
  \includegraphics[width=0.32\textwidth]{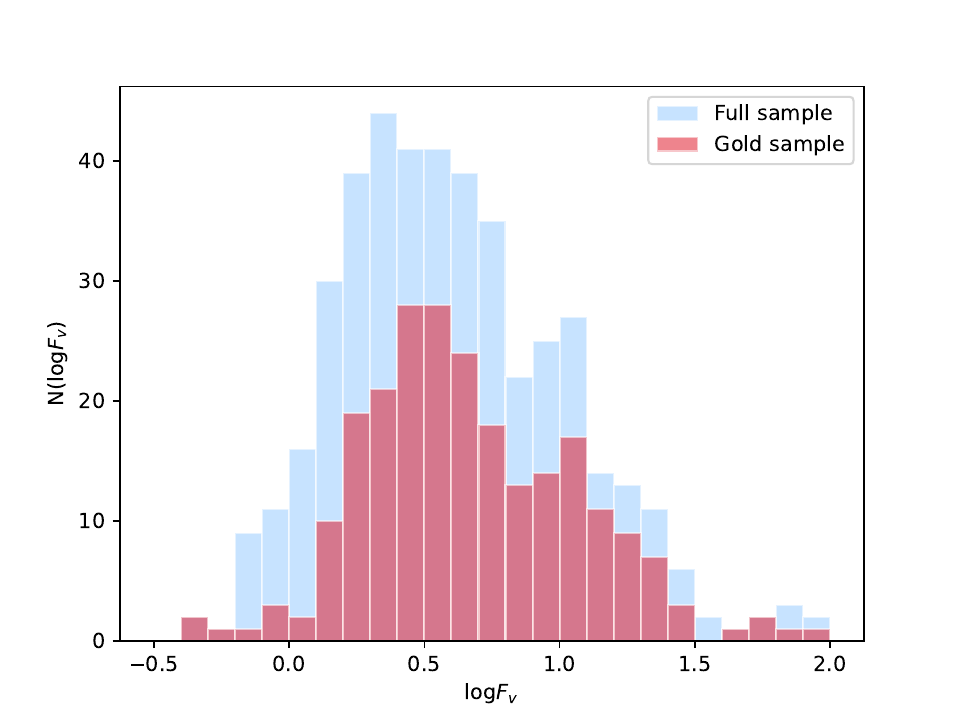}
  \includegraphics[width=0.32\textwidth]{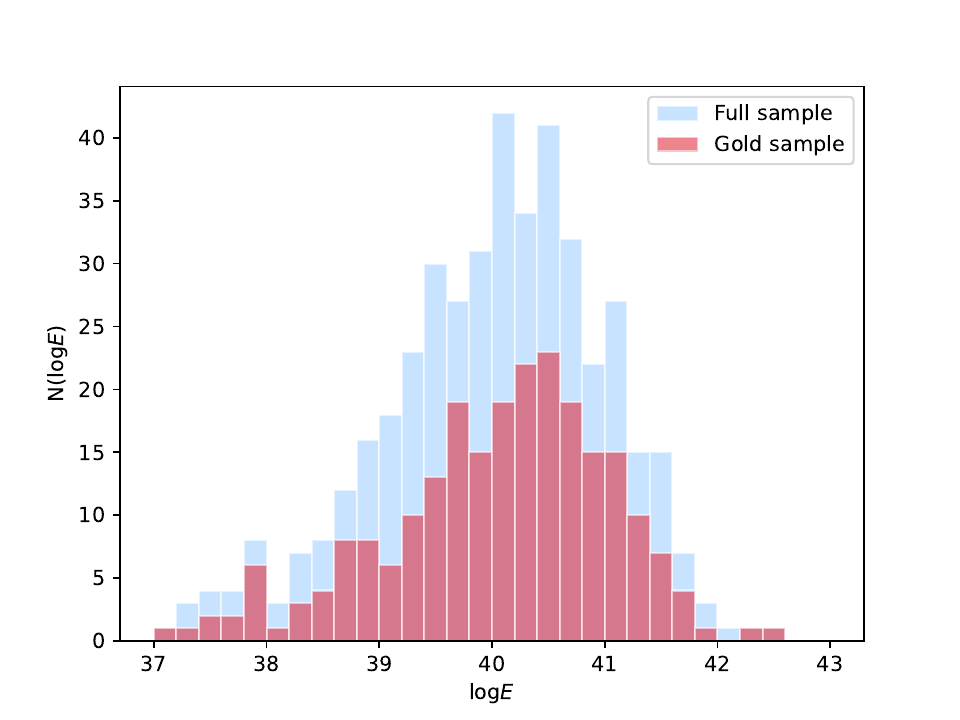}
  \includegraphics[width=0.32\textwidth]{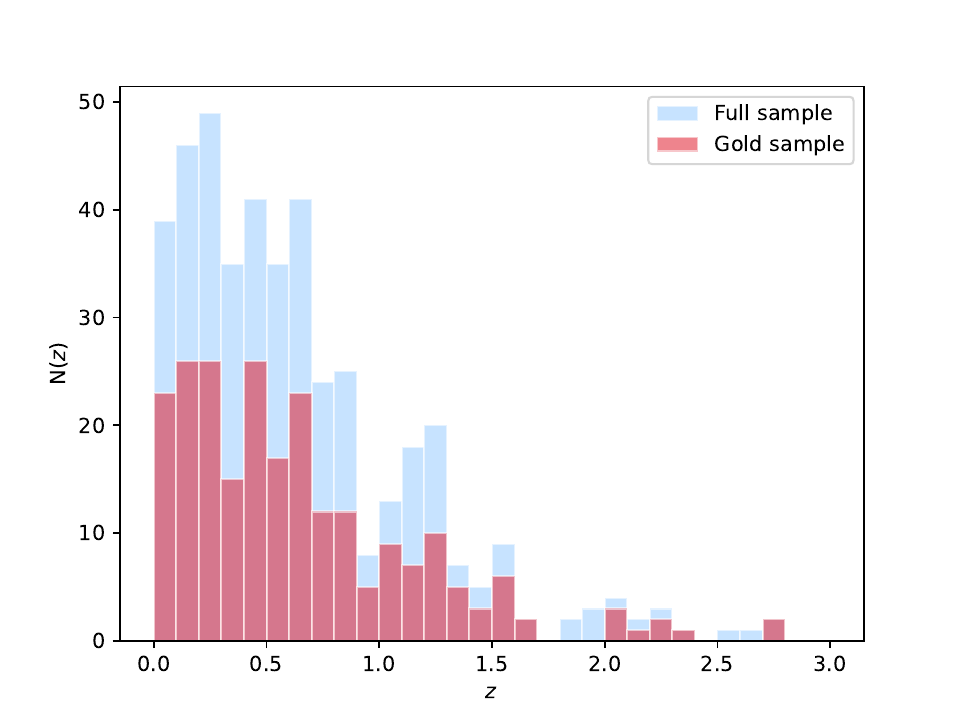}
  \caption{The histograms of fluence ($\mathrm{log}F_\nu$), isotropic energy ($\mathrm{log}E$) and inferred redshift ($z$) of the first non-repeating CHIME/FRB catalog.} \label{hist}
\end{figure}

Furthermore, we apply the following criteria put forward by \citet{CHIMEFRB:2021srp} to filter the data, in order to avoid bias in FRB population analyses: (1) $S/N > 12$; (2) ${\rm DM_{obs}} > 1.5{\rm max(DM_{NE2001}, DM_{YMW16})}$; (3) not detected in far sidelobes; (4) not detected during non-nominal telescope operations; (5) $\tau_{\rm scat}<10$ ms. The 236 FRBs filtered by the criteria are called the Gold sample, which will be used below to study the FRB population, and confront the results of using the Full sample. We uniformly divide redshift $z$ into 30 bins with stepsize $\Delta z=0.1$ in the range $(0, 3)$, divide energy $\mathrm{log}E$ into 30 bins with stepsize $\Delta {\rm log}E=0.2$ in the range $(37, 43)$, and divide fluence $\mathrm{log}F_\nu$ into 25 bins with stepsize $\Delta {\rm log}F_\nu=0.1$ in the range $(-0.5, 2)$. The histograms of these properties of the Full sample and Gold sample are shown in Figure \ref{hist}.

\section{The FRB population}\label{sec:population}

As now we have the distributions of fluence, energy and redshift, we can use them to constrain the FRB population. Several independent works regardless of the redshift distribution show that the intrinsic energy distribution roughly follows a power low $dN/dE \propto  E^{-\alpha}$, with $1.8 \lesssim \alpha \lesssim  2.0$ \citep{Lu:2019pdn,Lu:2020nsg,Luo:2018tiy,Luo:2020wfx}. It may have a high-energy exponential cutoff $E_c$ \citep{Luo:2020wfx,Lu:2020nsg}, which is not well constrained yet \citep{Zhang:2020ass}. Therefore, we model the intrinsic energy distribution as the cutoff power law,
\begin{equation}\label{eq:pE}
 p(E) \propto \left(\frac{E}{E_c}\right)^{-\alpha} \mathrm{exp}\left( -\frac{E}{E_c}\right).
\end{equation}

The redshift distribution in the observer frame can be derived from the intrinsic event-rate density $dN/dtdV$ according to the relation \citep{Zhang:2020ass,Zhang:2021kdu}
\begin{equation}\label{eq:pz}
  p(z) \propto \frac{dt}{dt_{\mathrm{obs}}}\frac{dN}{dtdV}\frac{dV}{dz},
\end{equation}
where $dt/dt_{\mathrm{obs}} = (1+z)^{-1}$ accounts for the cosmic expansion, and the redshift-dependent specific comoving volume $dV/dz$ in the flat $\Lambda \mathrm{CDM}$ model can be expressed as 
\begin{equation}
  \frac{dV}{dz} = \frac{4\pi cd_\mathrm{C}^2}{H_0 \sqrt{\Omega_m(1+z)^3 + \Omega_\Lambda}},
\end{equation}
where $d_\mathrm{C}$ is the comoving distance.

Actually, the intrinsic event-rate density $dN/dtdV$ is poorly constrained. The most well-motivated model is the SFH model. Several forms regarding the SFH have been proposed, e.g., \citet{Yuksel:2008cu} fits a wide range of SFH data with a three-segment empirical model, and a two-segment empirical model proposed in \citet{Madau:2014bja} is also widely adopted, soon an updated version of which is used in \cite{Madau:2016jbv}. However, no proof shows there is significant priority among these forms. For consistency, we follow \citet{Qiang:2021ljr} to adopt the updated version of two-segment empirical SFH model, i.e.,
\begin{equation}
  \mathrm{SFH} = \frac{(1+z)^{2.6}}{1+((1+z)/3.2)^{6.2}}.
\end{equation}

Several studies show that the intrinsic event-rate density doesn't simply track SFH \citep{Zhang:2021kdu,Qiang:2021ljr}. It is argued in \citet{Kulkarni:2018ola} that the research history of FRBs may resemble that of GRBs, which is found to follow an enhanced evolution with regard to SFH, i.e, $dN/dtdV \propto  (1+z)^{1.5} \times \mathrm{SFH}(z)$ \citep{Kistler:2007ud,Yuksel:2008cu}. This prompts us to take some similar SFH-related models into consideration. In this work, we reconsider the revised SFH-related redshift distribution models put forward by \citet{Qiang:2021ljr}. The models are enumerated below.
\begin{enumerate}[label=\textbullet]
  \item Power law (PL) model:
  \begin{equation}\label{eq:PL}
    \frac{dN}{dtdV} \propto  (1+z)^\gamma \times \mathrm{SFH}(z).
  \end{equation}
  \item Power law with an exponential cutoff (CPL) model:
  \begin{equation}
    \frac{dN}{dtdV} \propto (1+z)^\gamma \mathrm{exp}(-\frac{z}{z_c}) \times \mathrm{SFH}(z).
  \end{equation}
  \item SFH with an exponential cutoff (CSFH) model:
  \begin{equation}
    \frac{dN}{dtdV} \propto \mathrm{exp}(-\frac{z}{z_c}) \times \mathrm{SFH}(z).
  \end{equation}
  \item Two-segment (TSE) model:
  \begin{equation}
    \frac{dN}{dtdV} \propto \frac{(1+z)^{\gamma_1}}{1+((1+z)/s)^{\gamma_1 + \gamma_2}} \times \mathrm{SFH}(z).
  \end{equation}
  \item Non-SFH-based two-segment redshift distribution (TSRD) model:
  \begin{equation}\label{eq:TSRD}
    \frac{dN}{dtdV} \propto \frac{(1+z)^{a}}{1+((1+z)/C)^{a + b}}.
  \end{equation}
\end{enumerate}

We should note that equation (\ref{eq:pE}) and equation (\ref{eq:pz}) don't actually provide the energy and redshift distributions detected by the telescope. To transform them to the actually detected distributions, the selection effect should be considered. Unfortunately, the selection effect of the CHIME telescope is hard to determine. For simplicity, we take the selection model proposed by \citet{Zhang:2021kdu}. This model assumes the so-called `grey zone' between the minimum specific threshold fluence, i.e., $\mathrm{log}F_{\nu,\rm th}^\mathrm{min} = -0.5$ (as shown in the CHIME catalog), and the maximum specific threshold fluence $\mathrm{log}F_{\nu,\rm th}^\mathrm{max}$. In the `grey zone', FRBs are not fully detected. The detection efficiency is modeled by the power law form, 
\begin{equation}\label{eq:selection}
  \eta_\mathrm{det}(F_\nu) =
    \left(\frac{\mathrm{log}F_{\nu}-  \mathrm{log}F_{\nu,\rm th}^\mathrm{min}}{\mathrm{log}F_{\nu,\rm th}^\mathrm{max}-  \mathrm{log}F_{\nu,\rm th}^\mathrm{min}}\right)^n, ~~~ \mathrm{log}F_{\nu,\rm th}^\mathrm{min}< \mathrm{log}F_{\nu}< \mathrm{log}F_{\nu,\rm th}^\mathrm{max}.
\end{equation}
For $\mathrm{log}F_{\nu} \leq \mathrm{log}F_{\nu,\rm th}^\mathrm{min}$, FRBs are undetectable, hence $\eta_\mathrm{det}=0$; while for $\mathrm{log}F_{\nu} \geq \mathrm{log}F_{\nu,\rm th}^\mathrm{max}$, FRBs are fully deteceted, hence $\eta_\mathrm{det}=1$. In the paper of \citet{Zhang:2021kdu}, $n=3$ is fixed, and the parameter $\mathrm{log}F_{\nu,\rm th}^\mathrm{max}$ is manually adjusted to fit the observed fluence distribution. Here we treat both $n$ and $\mathrm{log}F_{\nu,\rm th}^\mathrm{max}$ as free parameters. The specific fluence $F_\nu$ can be written as a function of isotropic energy and redshift according to equation (\ref{eq:energy}), which reads
\begin{equation}\label{eq:fluence}
  F_\nu = \frac{(1+z)^{2+\beta}}{4\pi d_\mathrm{L}^2\Delta\nu} E.
\end{equation}

Since the redshift distribution and intrinsic energy distribution are independent with each other, the joint distribution of redshift and energy can be simply written as $p(z,E)=p(z)p(E)$. Then according to the probability theory, the detected fluence distribution after considering the selection effect can be written as
\begin{equation}\label{Fv_obs}
  p_\mathrm{det}(F_\nu) = \frac{d}{dF_\nu} \mathop{\iint }\limits_{F_\nu(E, z) \leq F_\nu} p(z)p(E) \eta_\mathrm{det}(F_\nu(E, z))  \,dE \,dz,
\end{equation}
where the 2-dimensional integral region is defined by $F_\nu(E, z) \leq F_\nu$, with $F_\nu(E, z)$ given by equation (\ref{eq:fluence}). Furthermore, the detected energy distribution can be written as
\begin{equation}\label{E_obs}
  p_\mathrm{det}(E) = \int_{z_\mathrm{min}}^{z_\mathrm{max}}p(z)p(E) \eta_\mathrm{det}(F_\nu(E, z))  \,dz.
\end{equation}
Similarly, the detected redshift distribution reads
\begin{equation}\label{z_obs}
  p_\mathrm{det}(z) = \int_{E_\mathrm{min}}^{E_\mathrm{max}}p(z)p(E) \eta_\mathrm{det}(F_\nu(E, z))  \,dE.
\end{equation}
We can further construct the cumulative distribution function (CDF) of each quantity,
\begin{equation}\label{eq:cdf}
    N_\mathrm{det}(>Q) = A\int_{Q}^{Q_\mathrm{max}}p_\mathrm{det}(Q) dQ, ~~~ Q = \left\{F_\nu,\, E,\, z\right\},
\end{equation}
where $Q$ is the quality that represents $F_\nu$, $E$ or $z$, and A is the normalization factor scaled to the total number of data points.

Equations (\ref{eq:cdf}) can be used to fit the observed distribution of fluence, energy and redshift of the CHIME/FRBs. In order to avoid the arbitrariness of the binning, the CDF rather than the differential distribution is used in the fitting. The $\chi^2$ of each distribution can be written as
\begin{equation}
  \chi^2_Q = \sum_{i}^{n} \frac{[N_\mathrm{det}(>Q_i) - N(>Q_i)]^2}{\sigma_{i}^2}, ~~~ Q = \left\{F_\nu,\, E,\, z\right\} ,
\end{equation}
where $N$ is the number of FRBs whose $F_\nu$, $E$ or $z$ is larger than that of the $i$-th FRB. The uncertainty is given by $\sigma = \sqrt{N}$ \citep{Wang:2019sio}. The total $\chi^2$ is constructed as
\begin{equation}\label{chi2total}
  \chi^2_{\mathrm{total}} = \chi^2_{F_\nu} + \chi^2_z + \chi^2_E,
\end{equation}
based on which the likelihood of observing a sample of FRBs with $F_\nu$, $E$ and $z$ can be written as
\begin{equation}\label{likelihood}
  \mathcal{L}(\mathrm{FRBs|parameters})  \propto \mathrm{exp}\left(-\frac{1}{2}\chi^2_{\mathrm{total}}(F_\nu,\, E,\, z)\right),
\end{equation}
where the parameters are power law index $\alpha$ and cutoff energy $E_c$ for energy distribution, $n$ and $\mathrm{log}F_{\nu,th}^\mathrm{max}$ for selection function, and the parameters in different redshift distribution models listed in equations (\ref{eq:PL}) -- (\ref{eq:TSRD}). Then the posterior probability distribution can be constructed according to Bayesian theorem,
\begin{equation}\label{posterior}
  P(\mathrm{parameters|FRBs})\propto\mathcal{L}(\mathrm{FRBs|parameters})P_0(\mathrm{parameters}),
\end{equation}
where $P_0$ is the prior of the parameters.

\begin{figure}[htbp]
  \centering
  \includegraphics[width=0.32\textwidth]{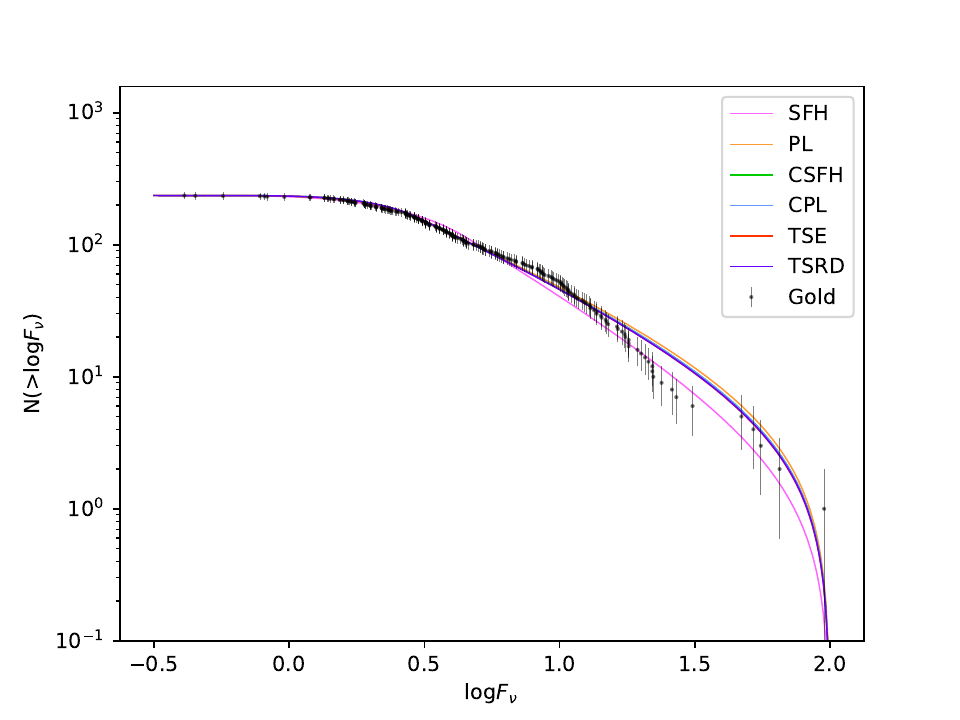}
  \includegraphics[width=0.32\textwidth]{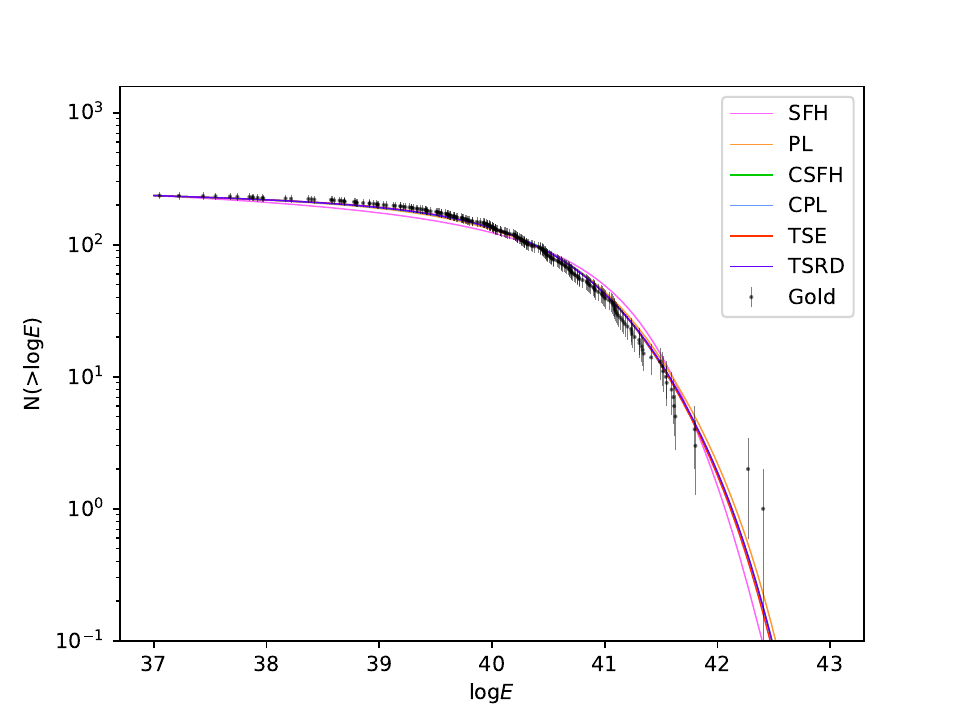}
  \includegraphics[width=0.32\textwidth]{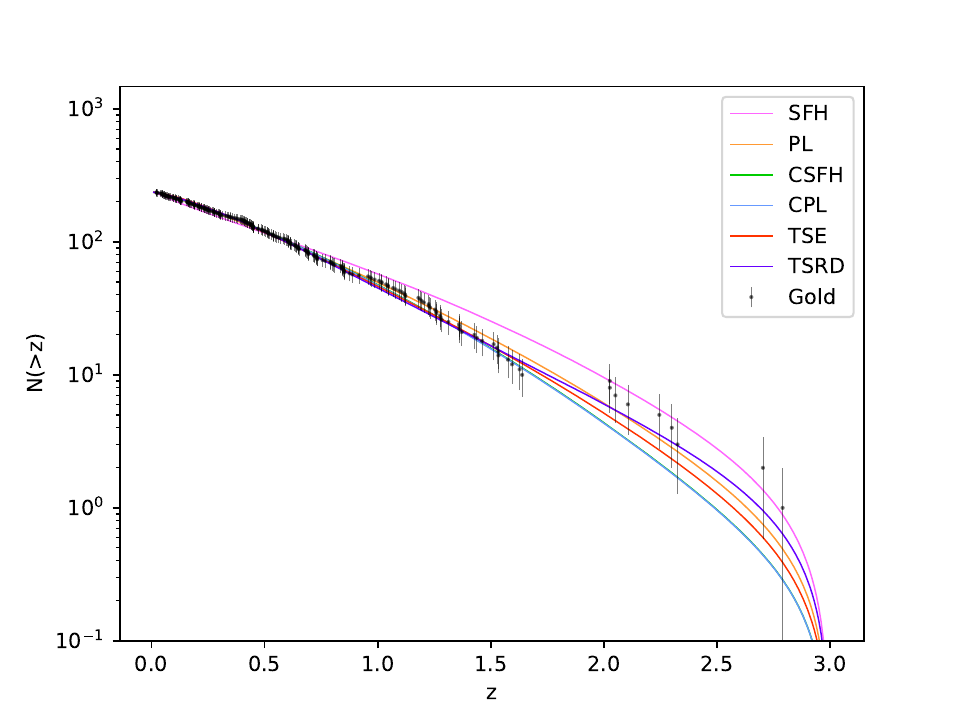}
  \includegraphics[width=0.32\textwidth]{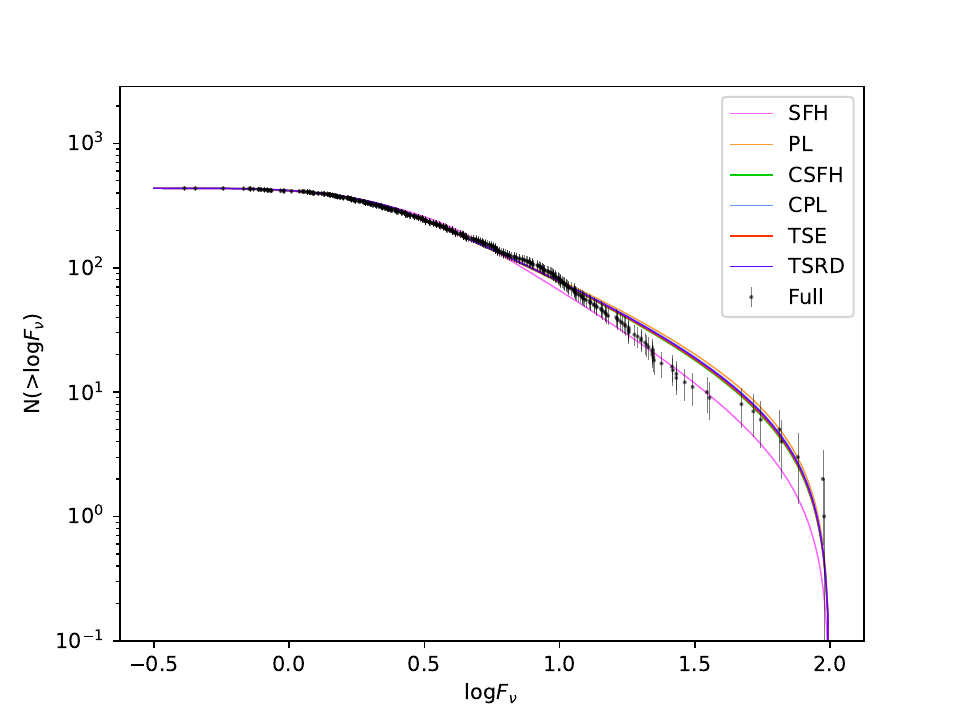}
  \includegraphics[width=0.32\textwidth]{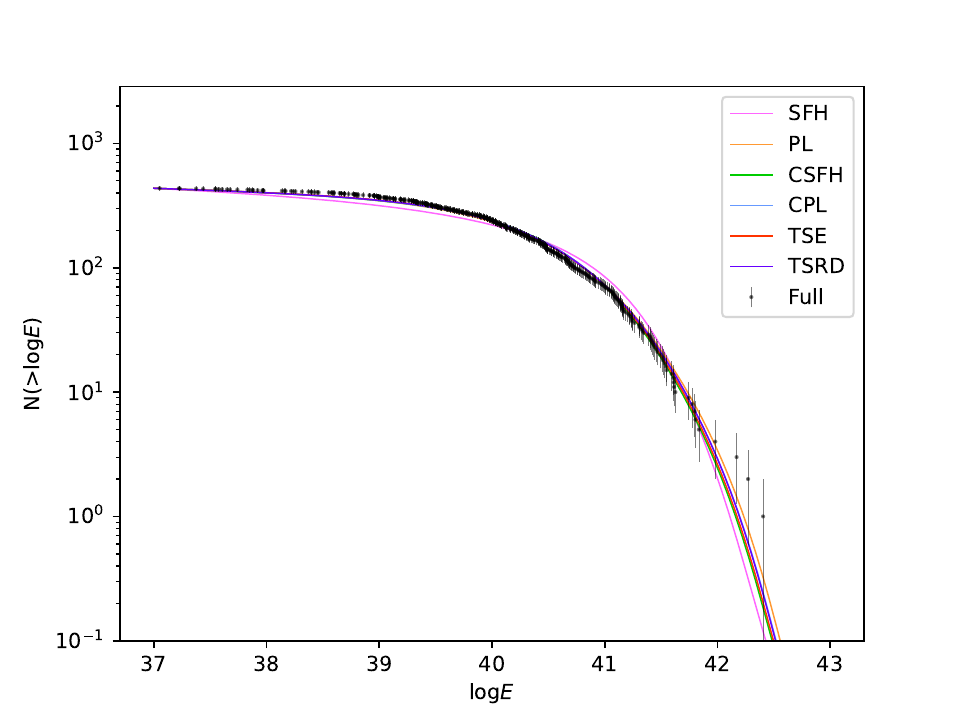}
  \includegraphics[width=0.32\textwidth]{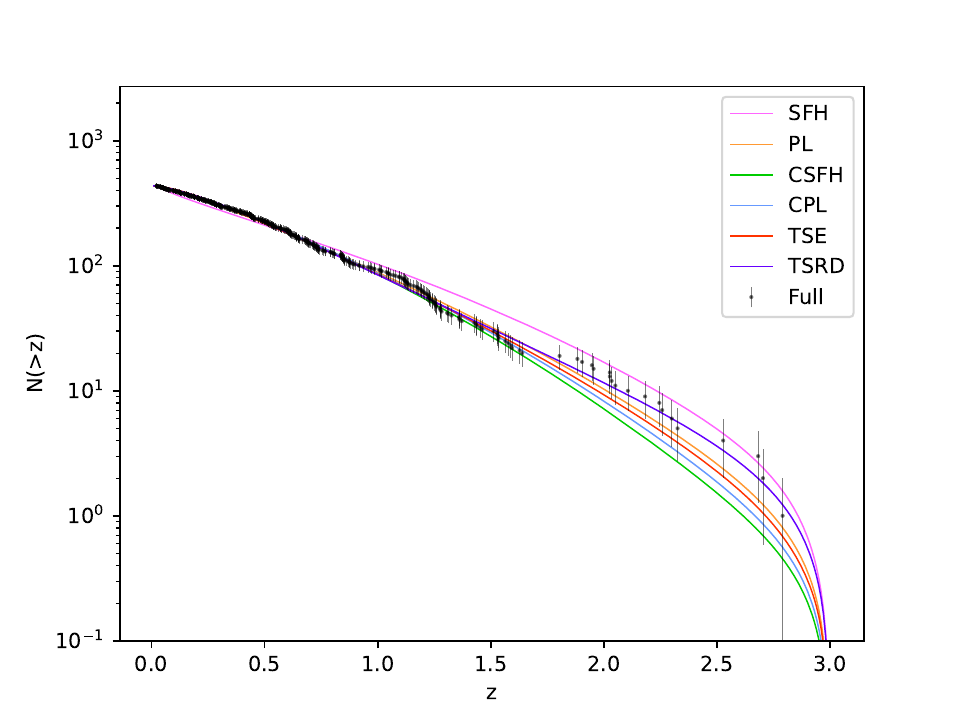}
  \caption{The best-fitting lines of the Gold sample (top panels) and the Full sample (bottom panels). The lines are plotted using equation (\ref{eq:cdf}) with the best-fitting parameters given in Table \ref{table:gold_sample} and Table \ref{table:full_sample}. Figures from left to right show the fits to fluence, energy and redshift, respectively.} \label{fitsplot}
\end{figure}

Through equation (\ref{posterior}), one can calculate the best-fitting parameters and their $1\sigma$ confidence level with the public Python package \textsf{emcee} \citep{Foreman-Mackey:2012any}. We take flat priors for all the parameters in a broad range, i.e., $\alpha \in U(1, 5)$, $\mathrm{log}E_c \in U(39, 44)$, $n \in U(1, 8)$, 
$\mathrm{log}F_{\nu,th}^\mathrm{max} \in (-0.49, 2.0)$, $\gamma, \gamma_1, \gamma_2, a, b \in U(-10, 10)$ and $z_c, s, C \in U(0.1, 10)$. The posterior PDFs of model parameters and the 2-dimensional confidence contours are shown in Figures \ref{fig:corner1}--\ref{fig:corner3} in the appendix. The median values and $1\sigma$ uncertainties of parameters are summarized in Table \ref{table:gold_sample} and Table \ref{table:full_sample} for the Gold sample and the Full sample, respectively. The best-fitting lines are depicted in Figure \ref{fitsplot}.

\begin{table}[htbp]
  \centering
  \caption{The best-fitting parameters and their $1\sigma$ confidence level by simultaneously constraining three sets of parameters using the Gold sample. Column 1: model name; Column 2-3: parameters in energy distribution; Column 4-5: parameters in selection function; Column 6-8: parameters in different redshift distribution models; Column 9: $\Delta\mathrm{BIC}$ with CSFH the fiducial model.}\label{table:gold_sample}
{\begin{tabular}{c|cc|cc|ccc|c}
  \hline
  Model & $\alpha$ & $\mathrm{log}E_c/\mathrm{erg}$  & $\mathrm{log}F_{\nu,\rm th}^{\mathrm{max}}$ & $n$ &  \multicolumn{3}{c|}{Model parameter}  & $\Delta\mathrm{BIC}$ \\
  \hline
  SFH & $2.13^{+0.01}_{-0.01}$& $42.03^{+0.01}_{-0.01}$& $0.66^{+0.01}_{-0.01}$& $4.45^{+0.05}_{-0.05}$& N/A & N/A  & N/A   & $555.15$\\
  PL& $1.83^{+0.01}_{-0.01}$& $42.07^{+0.01}_{-0.01}$& $0.46^{+0.01}_{-0.01}$& $5.27^{+0.19}_{-0.23}$& $\gamma = -2.16^{+0.06}_{-0.06}$& N/A &N/A   & $34.25$\\
  CSFH & $1.87^{+0.01}_{-0.01}$& $42.15^{+0.02}_{-0.02}$& $0.47^{+0.02}_{-0.01}$& $5.45^{+0.28}_{-0.33}$& $ z_c = 0.93^{+0.03}_{-0.03}$& N/A  &N/A   & $0$(fiducial)\\
  CPL  & $1.87^{+0.01}_{-0.01}$& $42.15^{+0.02}_{-0.02}$& $0.47^{+0.02}_{-0.01}$& $5.43^{+0.31}_{-0.33}$& $ \gamma =-0.04^{+0.30}_{-0.28}$& $z_c = 0.94^{+0.15}_{-0.11}$& N/A   & $6.57$\\
  TSE  & $1.89^{+0.01}_{-0.01}$& $42.13^{+0.02}_{-0.02}$& $0.49^{+0.01}_{-0.02}$& $5.24^{+0.37}_{-0.27}$& $ \gamma_1 =7.14^{+1.95}_{-2.15}$& $ \gamma_2 = 2.25^{+0.11}_{-0.09}$& $s = 1.05^{+0.02}_{-0.02}$& $2.25$\\
  TSRD   & $1.88^{+0.01}_{-0.01}$& $42.15^{+0.02}_{-0.02}$& $0.48^{+0.02}_{-0.01}$& $5.44^{+0.29}_{-0.31}$& $a =3.28^{+0.82}_{-0.65}$& $b = 1.35^{+0.39}_{-0.33}$& $C = 1.50^{+0.19}_{-0.14}$& $7.91$\\
  \hline
\end{tabular}}
\end{table}

\begin{table}[htbp]
  \centering
  \caption{The best-fitting parameters and their $1\sigma$ confidence level by fitting to the Full sample, given in the same manner as Table \ref{table:gold_sample}, but $\Delta\mathrm{BIC}$ is given with TSE the fiducial model.}\label{table:full_sample}
{\begin{tabular}{c|cc|cc|ccc|c}
  \hline
  Model & $\alpha$ & $\mathrm{log}E_c/\mathrm{erg}$  & $\mathrm{log}F_{\nu,\rm th}^{\mathrm{max}}$ & $n$ &  \multicolumn{3}{c|}{Model parameter} & $\Delta\mathrm{BIC}$ \\
  \hline
  SFH & $2.14^{+0.01}_{-0.01}$ & $41.96^{+0.01}_{-0.01}$ & $0.70^{+0.01}_{-0.01}$ & $3.53^{+0.02}_{-0.02}$ & N/A & N/A  & N/A  & $1997.01$\\
  PL&  $1.78^{+0.01}_{-0.01}$ & $42.12^{+0.01}_{-0.01}$ & $0.42^{+0.01}_{-0.01}$ & $3.79^{+0.07}_{-0.07}$  & $\gamma = -2.53^{+0.03}_{-0.03}$& N/A &N/A   & $75.91$\\
  CSFH & $1.84^{+0.01}_{-0.01}$ & $42.04^{+0.01}_{-0.01}$ & $0.49^{+0.01}_{-0.01}$ & $3.46^{+0.07}_{-0.05}$  & $ z_c = 0.84^{+0.02}_{-0.01}$ & N/A  &N/A   & $39.38$\\
  CPL  & $1.81^{+0.01}_{-0.01}$ & $42.08^{+0.01}_{-0.01}$ & $0.46^{+0.01}_{-0.01}$ & $3.57^{+0.06}_{-0.07}$  & $ \gamma =-0.99^{+0.15}_{-0.16}$& $z_c = 1.35^{+0.16}_{-0.12}$ & N/A  & $25.12$\\
  TSE  & $1.83^{+0.01}_{-0.01}$ & $42.05^{+0.01}_{-0.01}$ & $0.48^{+0.01}_{-0.01}$ & $3.50^{+0.10}_{-0.08}$  & $ \gamma_1 =9.13^{+0.65}_{-1.15}$ & $ \gamma_2 = 2.49^{+0.05}_{-0.05}$ & $s = 1.03^{+0.01}_{-0.01}$& $0$(fiducial)\\
  TSRD   & $1.82^{+0.01}_{-0.01}$ & $42.07^{+0.01}_{-0.01}$ & $0.47^{+0.01}_{-0.01}$ & $3.58^{+0.08}_{-0.07}$  & $a =3.06^{+0.54}_{-0.53}$ & $b = 1.39^{+0.22}_{-0.17}$ & $C = 1.39^{+0.12}_{-0.08}$ & $2.54$\\
  \hline
\end{tabular}}
\end{table}

To quantify the goodness of fit and to choose the best model, we calculate the Bayesian information criterion (BIC), which is defined by \citep{schwarz1978}
\begin{equation}
    \mathrm{BIC} = -2\mathrm{ln}\mathcal{L} _\mathrm{max} + k\mathrm{ln}(N),
\end{equation}
where $\mathcal{L} _\mathrm{max}$ is the maximum likelihood calculated at the best-fitting parameters, $k$ is the number of free parameters and $N$ is the number of data points. The model with a smaller value of BIC reflects a better fit. Note that it's not the absolute but the relative value of BIC that matters. Given the fiducial model, the relative value of  BIC is defined as
\begin{equation}\label{eq:dic}
    \Delta \mathrm{BIC_{model}} = \mathrm{BIC_{model}} - \mathrm{BIC_{fiducial}}.
\end{equation} 
According to the Jeffreys' scale \citep{jeffreys1998theory,liddle2007information}, $\Delta \mathrm{BIC} \geq 5$ or $\Delta \mathrm{BIC} \geq 10$ indicates that there is `strong' or `decisive' evidence against the considered model with respect to the fiducial model. For the Gold sample, we take CSFH as the fiducial model, for it has the smallest BIC. For the Full sample, we take TSE as the fiducial model for the same reason. We list the $\Delta \mathrm{BIC}$ values for each model in the last column of Table \ref{table:gold_sample} and Table \ref{table:full_sample}.

With the results listed in Table \ref{table:gold_sample} and Table \ref{table:full_sample}, we have the rank of fits $\mathrm{CSFH\approx TSE > CPL\approx TSRD>PL\gg SFH}$ for the Gold sample, and $\mathrm{TSE\approx TSRD>CPL>CSFH>PL\gg SFH}$ for the full sample. According to the $\Delta \mathrm{BIC}$ values, both samples strongly disfavour the SFH model, compared with the rest five models. However, distinct difference exists between Gold sample and Full sample. For example, CSFH model is the best model for the Gold sample, while it is the third worst model for the Full sample. In addition, the selection function parameter $n$ differs significantly between Gold sample and Full sample.

\begin{table}
  \centering
  \caption{The $p_\mathrm{ks}$ values for $\mathrm{log}F_\nu$, $\mathrm{log}E$ and $z$ of different redshift models for the Gold sample.}\label{table:ks_gold}
  {\begin{tabular}{cccc}
    \hline
    Model&$p_\mathrm{ks}$ for $\mathrm{log}F_\nu$ & $p_\mathrm{ks}$ for $\mathrm{log}E$ & $p_\mathrm{ks}$ for $z$\\
    \hline
    SFH&$0.2009$&$0.6014$&$0.3561$\\
    PL&$0.8270$&$0.2345$&$0.9124$\\
    CSFH&$0.7108$&$0.3755$&$0.9706$\\
    CPL&$0.5685$&$0.3911$&$0.9648$\\
    TSE&$0.6566$&$0.4647$&$0.6509$\\
    TSRD&$0.6250$&$0.3800$&$0.8563$\\
    \hline
  \end{tabular}}
\end{table}

\begin{table}
  \centering
  \caption{The $p_\mathrm{ks}$ values for $\mathrm{log}F_\nu$, $\mathrm{log}E$ and $z$ of different redshift models for the Full sample.}\label{table:ks_full}
  {\begin{tabular}{cccc}
    \hline
    Model&$p_\mathrm{ks}$ for $\mathrm{log}F_\nu$ & $p_\mathrm{ks}$ for $\mathrm{log}E$ & $p_\mathrm{ks}$ for $z$\\
    \hline
    SFH&$0.1879$&$3.58\times10^{-8}$&$0.0189$\\
    PL&$0.5886$&$0.0190$&$0.6780$\\
    CSFH&$0.6781$&$0.0239$&$0.7282$\\
    CPL&$0.7542$&$0.0309$&$0.9011$\\
    TSE&$0.7027$&$0.0614$&$0.5358$\\
    TSRD&$0.7656$&$0.0341$&$0.6445$\\
    \hline
  \end{tabular}}
\end{table}

Furthermore, we perform the Kolmogorov-Smirnov (KS) test in the same manner as in \citet{Qiang:2021ljr}, to check if the best-fitting models can pass the statistical test or not. To this end, we confront the data points with the best-fitting CDF curve of each model, and calculate the p-values ($p_\mathrm{ks}$), which are summarized in Table \ref{table:ks_gold} for the Gold sample and Table \ref{table:ks_full} for the Full sample. The null hypothesis that the data points are drawn from the model distribution can not be rejected if $p_\mathrm{ks}>0.1~\mathrm{or}~0.05$. As can be seen, for the Gold sample, all models (even the worst model SFH) successfully pass the KS test, although some models have smaller $p_\mathrm{ks}$ values than that of \citet{Qiang:2021ljr}. However, the situation is different for the Full sample. All models except for SFH successfully pass the KS test for fluence and redshift distributions, but none of the model can pass the KS test for energy distribution. This is because in the Bayesian framework, the parameters are optimized by minimizing the total $\chi_{\rm total }^2$ (equation \ref{chi2total}), rather than the individual $\chi^2(F_\nu)$, $\chi^2(E)$ or $\chi^2(z)$, which may induce bias for some observables. In this case, the KS test may be conflicted with the results inferred from the Bayesian method.

\section{Discussion and conclusions}\label{sec:conclusions}

In this paper, we investigate the FRB population based on the first CHIME/FRB catalog using Bayesian inference method. We first reconstruct the ${\rm DM_E}-z$ relation from 17 well-localized FRBs, then use it to infer the redshift of the CHIME/FRBs. Together with the observed fluence, the isotropic energy of each FRB is calculated. Thus, each FRB contains three observables: the redshift $z$, the fluence $F_\nu$, and the isotropic energy $E$. Finally, we construct a Bayesian framework to constrain the FRB population using the first CHIME/FRB catalog. We apply a set of criteria to exclude FRBs that may bias the population analysis, and call the remaining FRBs the Gold sample. Three factors can affect the observables of FRBs: the intrinsic energy distribution, the selection effect of detector, and the intrinsic redshift distribution. The intrinsic energy distribution is modeled by the power law with an exponential cutoff. The selection effect is modeled as a two-parametric function of fluence. The intrinsic redshift distribution is the factor that we are most interested in. The SFH model and five SFH-related redshift distribution models proposed by \citet{Qiang:2021ljr} are considered here. We construct a joint likelihood of fleunce, energy and redshift. All the free parameters are constrained simultaneously using Bayesian inference method. The best-fitting parameters are summarized in Table \ref{table:gold_sample} and Table \ref{table:full_sample} for the Gold sample and Full sample, respectively.

From Table \ref{table:gold_sample} and Table \ref{table:full_sample}, one can see that the energy distribution parameters ($\alpha$ and $\log E_c$) are tightly constrained. These two parameters are very stable, that is, they are almost independent of the models and data samples. Except for the SFH model, which has a power-law index $\alpha \approx 2.1$, the indices of all the rest five models fall into the range $1.8 \lesssim \alpha \lesssim 1.9$, well consistent with the previous findings \citep{Lu:2019pdn,Lu:2020nsg,Luo:2018tiy,Luo:2020wfx}. For the cutoff energy $E_c$, it is tightly constrained to be about $\mathrm{log}E_c\approx 42$ for both Full sample and Gold sample, which is slightly larger than the value ($\mathrm{log}E_c\approx 41$ or 41.5) used in \citet{Zhang:2021kdu} and \citet{Qiang:2021ljr}.

The selection function parameters ($\mathrm{log}F_{\nu,\rm th}^{\mathrm{max}}$ and $n$) are also tightly constrained. For the Gold sample, these two parameters are very stable, with $\mathrm{log}F_{\nu,\rm th}^{\mathrm{max}}\approx 0.47$, and $n\approx5.2\sim5.5$. However, the SFH model has a larger value of $\mathrm{log}F_{\nu,\rm th}^{\mathrm{max}}\approx 0.66$ and a smaller value of $n\approx4.45$. The value of $n$ we obtained here is about twice of that used by \citet{Zhang:2021kdu} and \citet{Qiang:2021ljr}. This is because \citet{Zhang:2021kdu} and \citet{Qiang:2021ljr} fixed $n=3$ based on the full sample. As we can see from Table \ref{table:full_sample}, the $n$ value constrained from the Full sample is in the range of $3.4 \lesssim n \lesssim 3.8$, consistent with the value used by \citet{Zhang:2021kdu} and \citet{Qiang:2021ljr}. The best-fitting value of $n$ differs significantly between Gold sample and Full sample. As is shown in the first histogram of Figure \ref{hist}, the Gold sample shows a steeper incline at low fluence and peaks slightly later than the Full sample, which can be properly translated to a larger $n$ and a slightly larger average $\mathrm{log}F_{\nu,\rm th}^\mathrm{max}$, in accordance with the best-fitting results. \citet{Qiang:2021ljr} fixed $\mathrm{log}E_c= 41.5$ and obtained the maximum specific threshold fluence $\mathrm{log}F_{\nu,\rm th}^\mathrm{max} \sim 0.75$, here we get a smaller $\mathrm{log}F_{\nu,\rm th}^\mathrm{max}$ but a larger $\mathrm{log}E_c$. The reason is that a smaller $\mathrm{log}F_{\nu,\rm th}^\mathrm{max}$ leads to a larger average fluence, which also requires a larger $E_c$ to match the distribution.

As for the redshift distribution models, the one-parametric PL model and CSFH model can be tightly constrained. However, some parameters in multi-parametric models have large uncertainties, such as $\gamma$ in CPL model, $\gamma_1$ in TSE model and $a$ in TSRD model. It is good to mention that we get results quite different from that in \citet{Qiang:2021ljr} where these models are initially proposed. For the one-parametric PL model and CSFH model, we get $\gamma=-2.16^{+0.06}_{-0.06}$ and $z_c = 0.93^{+0.03}_{-0.03}$ from the Gold sample, while \citet{Qiang:2021ljr} prefers larger values of $\gamma\approx -1.1$ and $z_c \approx 2.8$. For the two-parametric CPL model, we have $\gamma =-0.04^{+0.30}_{-0.28}$ and $z_c = 0.94^{+0.15}_{-0.11}$ from the Gold sample, in contrast to $\gamma \approx -0.6$ and $z_c \approx 5.5$ in \citet{Qiang:2021ljr}. For the three-parametric TSE model and TSRD model, the distinction between our work and \citet{Qiang:2021ljr} is even larger, though the uncertainty becomes larger accordingly.

Several reasons may play important roles in these differences. First, we infer redshifts of CHIME/FRBs with the distribution of $\mathrm{DM_{IGM}}$ and $\mathrm{DM_{host}}$ properly considered. In contrast, \citet{Qiang:2021ljr} assumes that $\mathrm{DM_{host}}$ is a constant and $\mathrm{DM_{IGM}}$ is given by the mean value, then calculate redshift by numerically solving equation (\ref{eq:DM_obs}). The method of \citet{Qiang:2021ljr} may cause significant bias on the redshift inference, and may lead to negative redshift for FRBs with small observed DM. Note that our method may still introduce some uncertainties, especially at high-redshift region. This is because the ${\rm DM_E}-z$ relation derived form the low-redshift data may be broken at high-redshift region. Nevertheless, there is no strong evidence for the redshift evolution of host DM within the present data \citet{Tang:2023qbg}. Therefore, the ${\rm DM_E}-z$ relation is still the best way to estimate the redshift of unlocalized FRBs. With more localized FRBs in the future, this problem could be alleviated. Second, the 452 CHIME/FRBs they used include both repeaters and nonrepeaters. In contrast, we only consider the non-repeaters with ${\rm DM_E}>100~{\rm pc~cm^{-3}}$. The repeaters and nonrepeaters may have different populations, and the redshift estimation of FRBs with small ${\rm DM_E}$ may be significantly biased. Third, \citet{Qiang:2021ljr} fixed $E_c$, $\alpha$ and $n$ to reduce freedom, and generated a large amount of mock FRBs at each grid in parameter space to confront the real sample by performing KS test, which not only consumes computational power but also reduces precision. In contrast, we construct the joint distribution of fluence, energy and redshift analytically, thus all the parameters can be constrained simultaneously using Bayesian inference method. Last but not least, they take priority in fitting the $N \propto F_\nu ^{-3/2}$ relation, but it might be more reasonable to equally weight the fluence, energy and redshift (or DM) distribution.

We also perform the KS test to check if the besting-fitting models can match the data points or not. We find that all models can successfully pass the KS test for the Gold sample. For the Full sample, however, none of the model can simultaneously pass the KS test for fluence, energy and redshift distributions. It is possible to adjust the model parameters such that all the three observables can simultaneously pass the KS test, but this compromises the status of these parameters as ``best-fitting''. The Bayesian method can provide us the ``best-fitting" parameters, but without commitment that all the three observables simultaneously pass the KS test. On the contrary, the KS test method adopted by \cite{Zhang:2021kdu} and \cite{Qiang:2021ljr} can manually adjust the parameters to ensure that all the three observables simultaneously satisfy the KS test, but have difficulty in finding the ``best-fitting" parameters. In addition, the KS test can't deal with the uncertainty of observations, and model selection based on KS test may lead to contradiction. For instance, based on the $p_\mathrm{ks}$ values of the Gold sample (Table \ref{table:ks_gold}), the PL model seems to fit the fluence distribution best, while the energy distribution prefers the SFH model, but the CSFH model match the redshift distribution best. Therefore, we use BIC rather than the KS test to judge the goodness of fit.

Above all, the non-parametric SFH model is strongly disfavoured by both the Gold sample and Full sample, as is shown by the $\Delta {\rm BIC}$ values in Table \ref{table:gold_sample} and \ref{table:full_sample}. According to BIC, the CSFH model performs the best fit to the Gold sample, while the TSE model fits the Full sample best. However, other models can also fit the data well, as long as the selection function of detector and parameters are properly chosen. For example, TSE model and CSFH model match the Gold sample equally well, while TSRD model and TSE model match the Full sample equally well, see the $\Delta {\rm BIC}$ values in Table \ref{table:gold_sample} and Table \ref{table:full_sample}. Therefore, we summarize that it is still premature to make a conclusive conclusion about the FRB population with the present data, but the hypothesis that FRBs follow the SFH can be definitely excluded. We hope that a large amount of well-localized FRBs observed in the near future can unambiguously determine the FRB population.

\begin{acknowledgements}
This work has been supported by the National Natural Science Fund of China under grant nos. 12147102, 12275034 and 12347101, and the Fundamental
Research Funds for the Central Universities of China under grant no. 2023CDJXY-048.
\end{acknowledgements}

\bibliography{reference}{}
\bibliographystyle{aasjournal}

\appendix
\section{THE CONTOUR PLOTS}
   
\begin{figure}[htbp]
    \centering
    \includegraphics[width=0.48\textwidth]{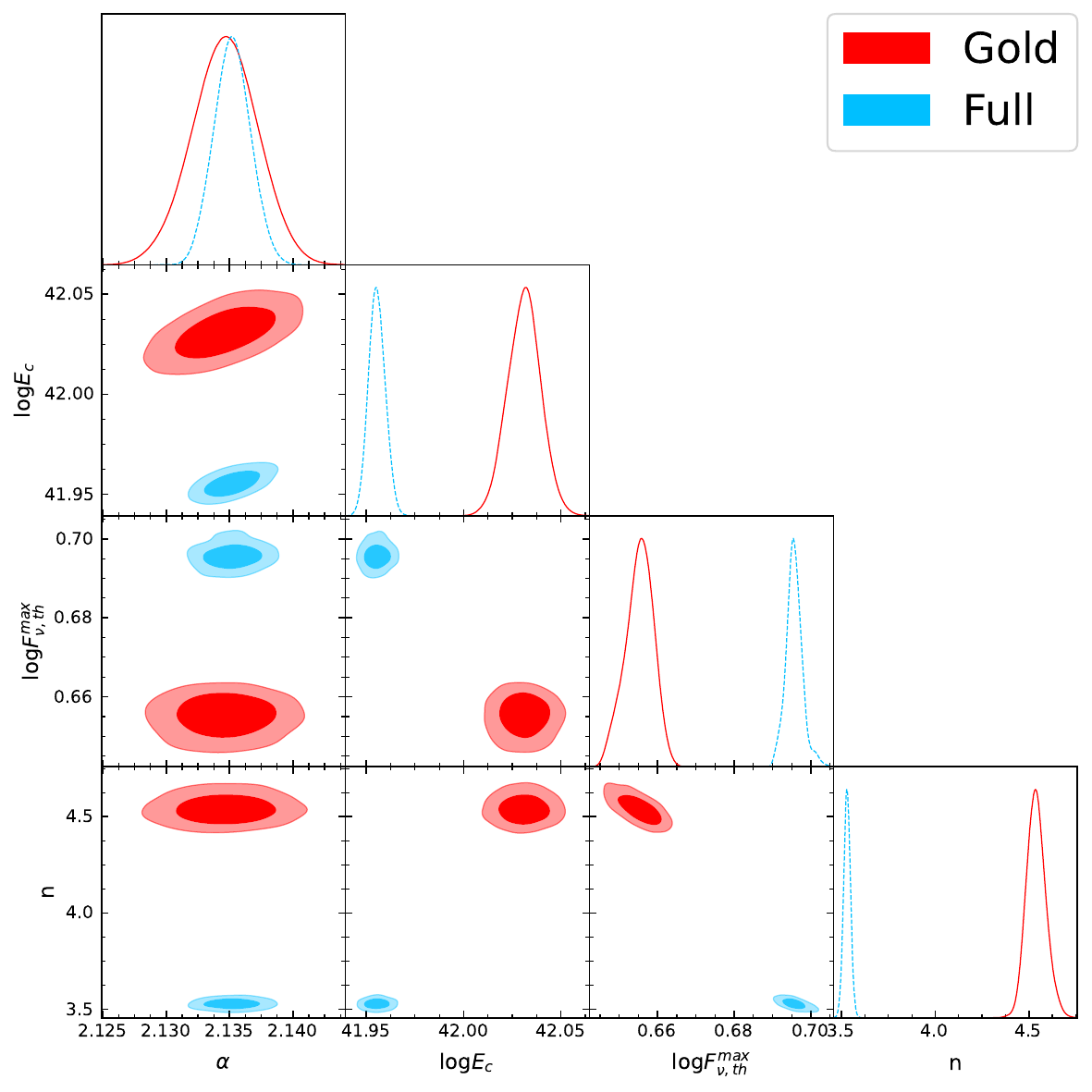}
    \includegraphics[width=0.48\textwidth]{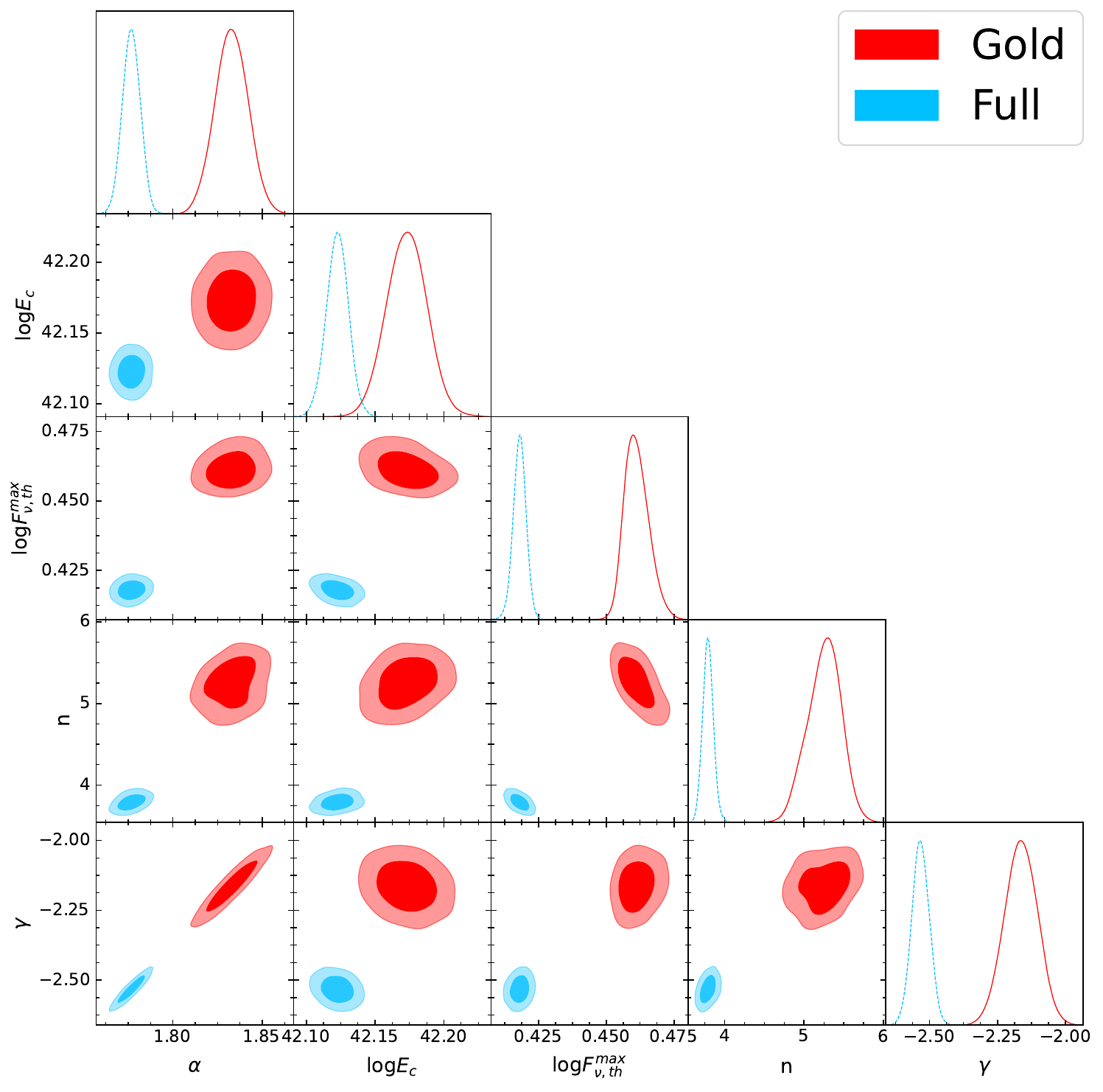} 
    \caption{The contour plot for the SFH model (left) and PL model (right).} \label{fig:corner1}
\end{figure}

\begin{figure}[htbp]
    \centering
    \includegraphics[width=0.48\textwidth]{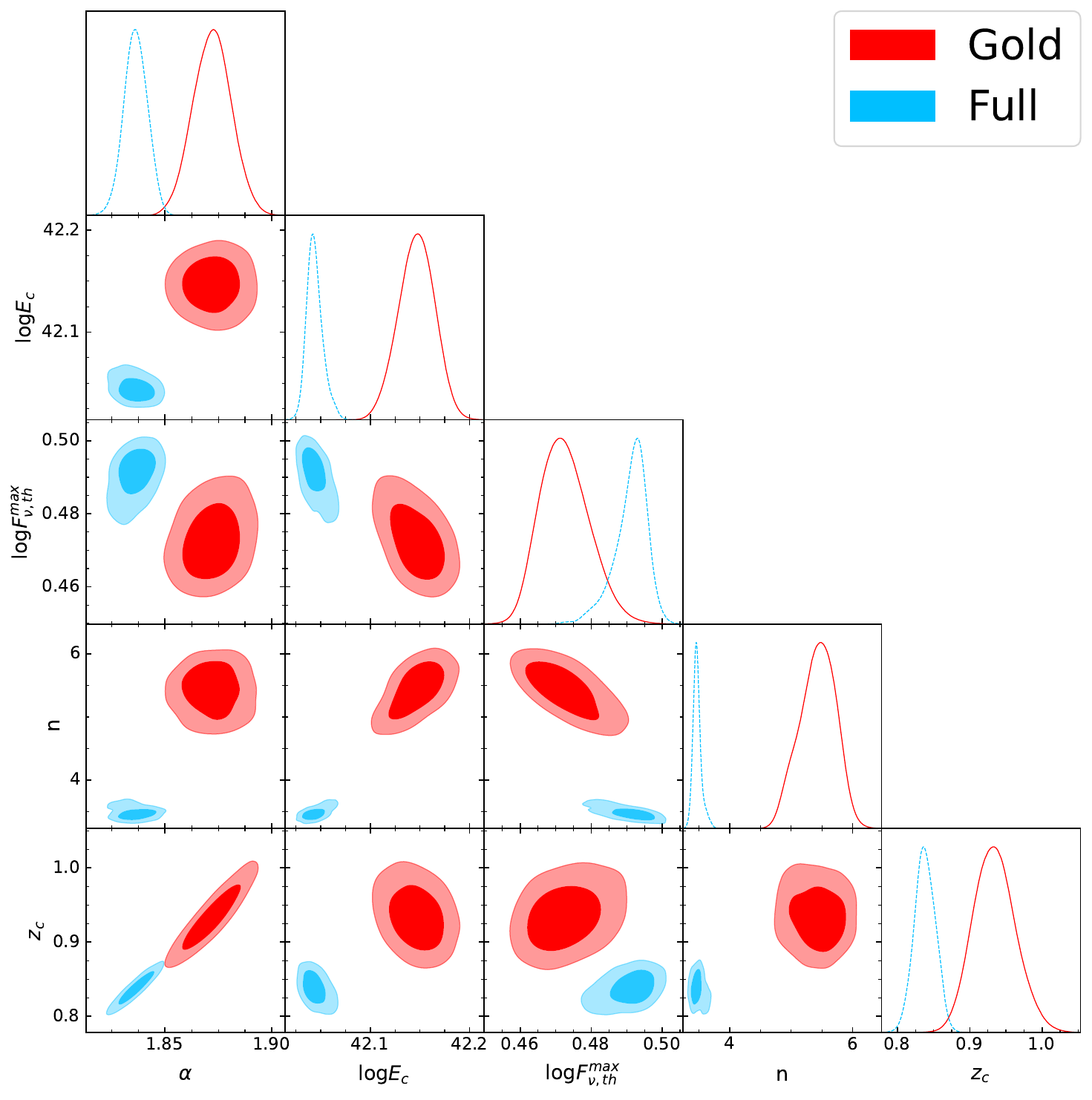}
    \includegraphics[width=0.48\textwidth]{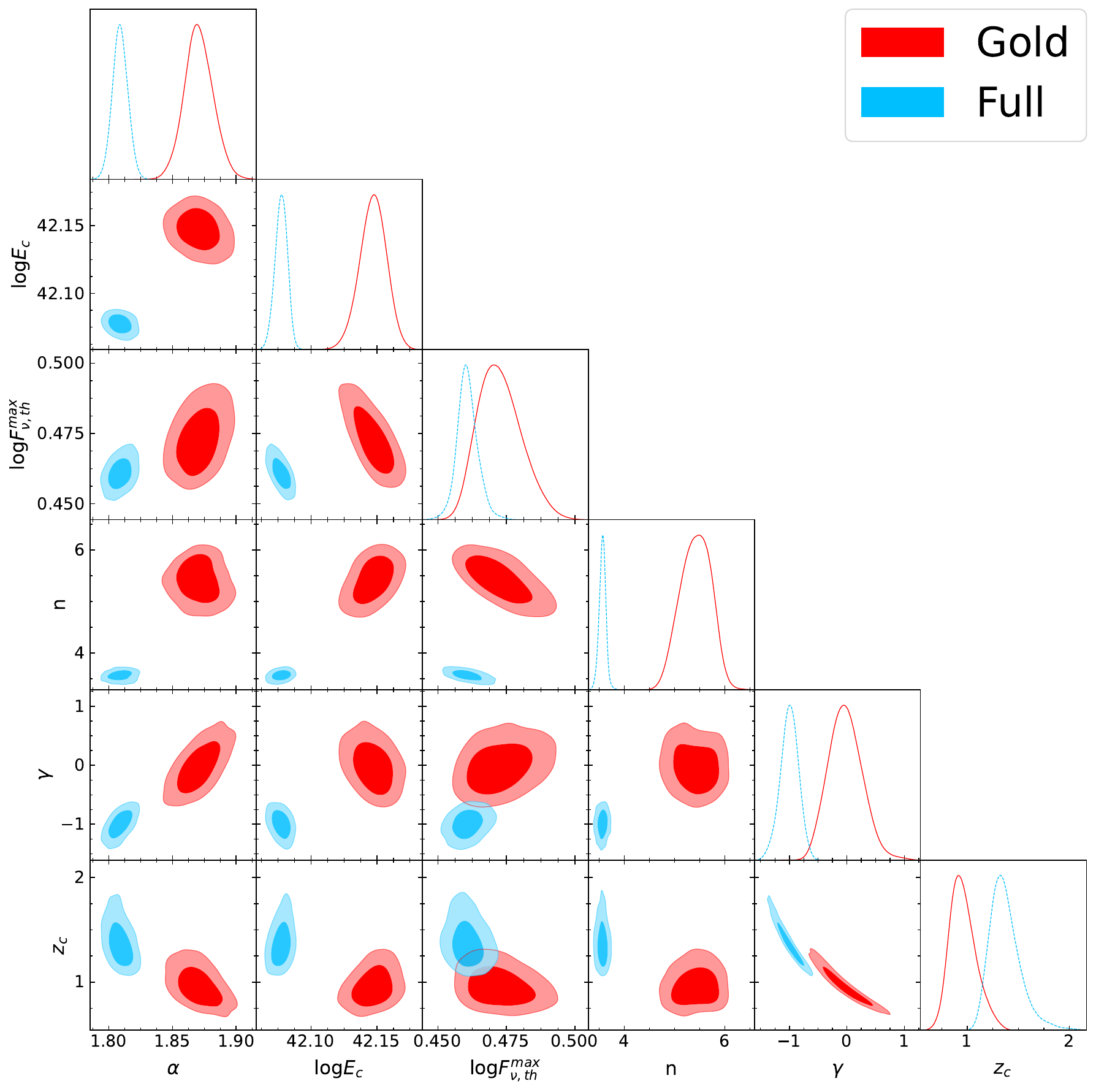}
    \caption{The contour plot for the CSFH model (left) and CPL model (right).} \label{fig:corner2}
\end{figure}

\begin{figure}[htbp]
    \centering
    \includegraphics[width=0.48\textwidth]{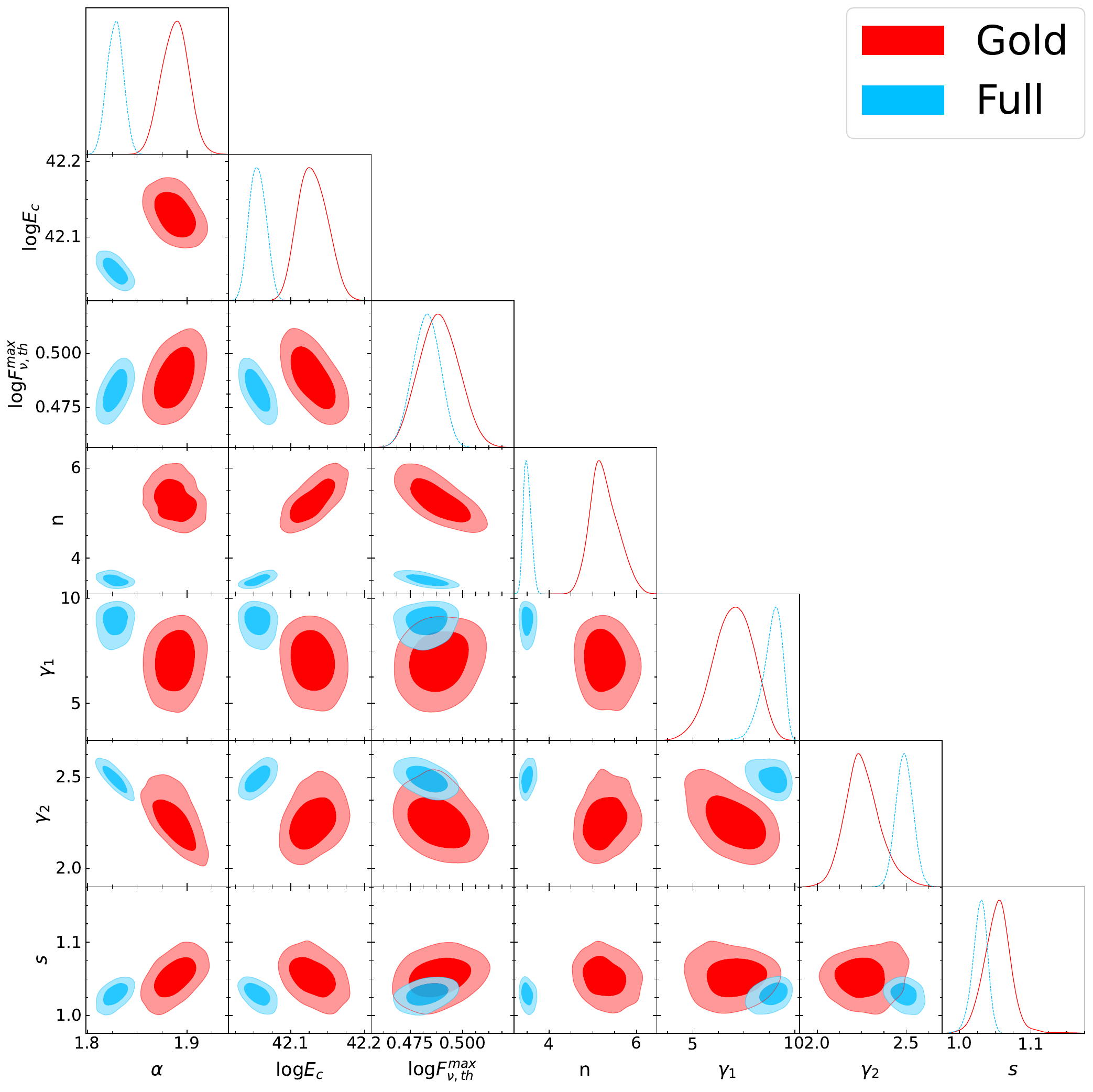}
    \includegraphics[width=0.48\textwidth]{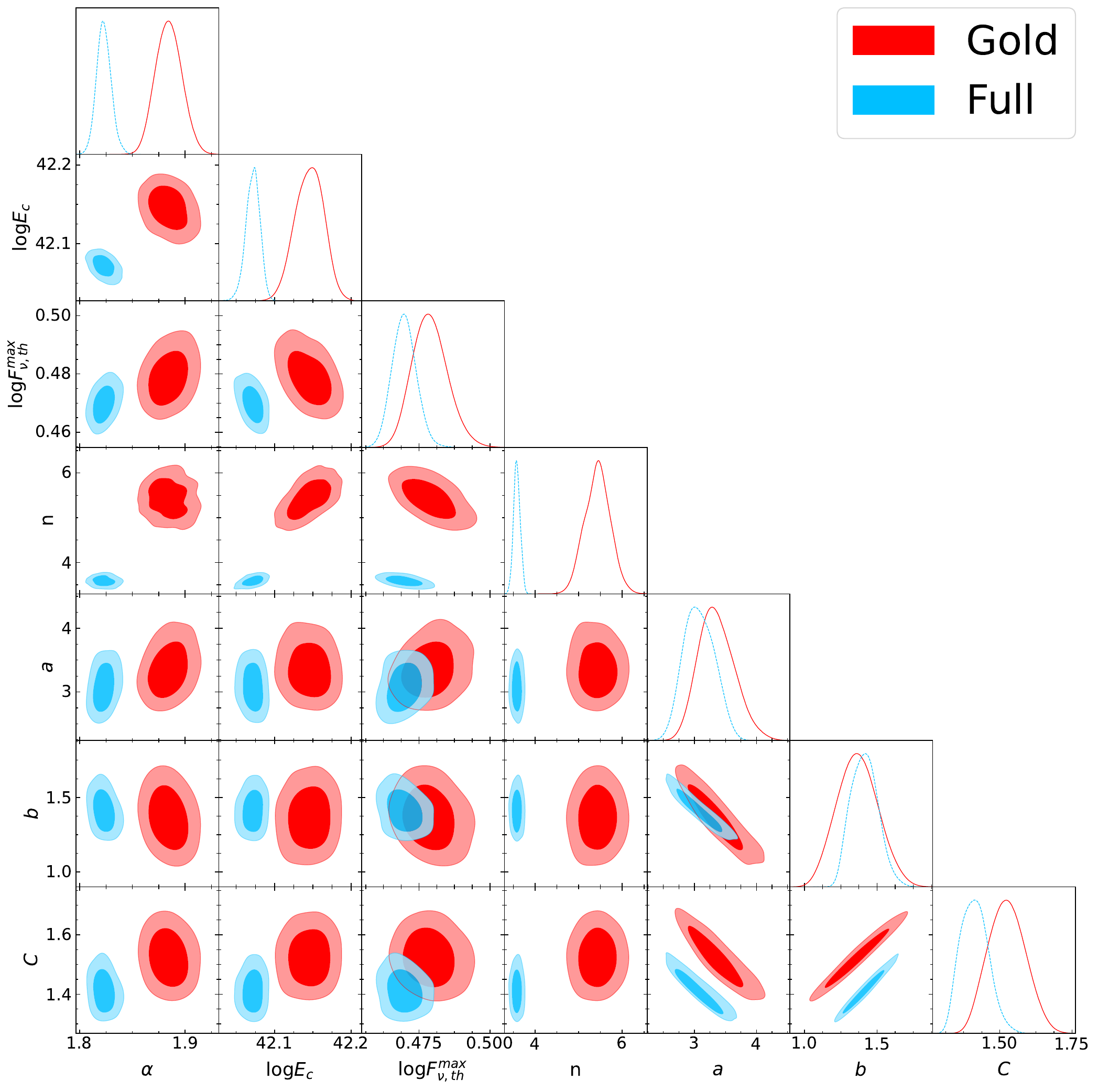}
    \caption{The contour plot for the TSE model (left) and TSRD model (right).} \label{fig:corner3}
\end{figure}

The contour plots for the SFH, PL, CSFH, CPL, TSE and TSRD models are shown in Figures (\ref{fig:corner1})--(\ref{fig:corner3}), respectively. For a specific model, the constraints from Full sample and Gold sample are plotted together for comparison.

\label{fig:contour2}

\label{fig:contour3}

\end{document}